\documentclass[twoside]{article}
\usepackage{a4wide}
\usepackage{amsmath}
\usepackage{amsfonts}
\usepackage{amssymb}
\usepackage{times} 
\def\date{Apr 5, 2005}
\newcommand{\ed}{\end{document}}
\renewcommand{\theequation}{\arabic{section}-\arabic{equation}}
\newcounter{mycnt}[section]

\def\!{\kern -0.15ex}

\begin{document}
\title{Born reciprocity and the granularity of space-time}
\author{P D Jarvis and S O Morgan}
%%%%if used by the author(s):
%{ {\renewcommand{\thefootnote}{\fnsymbol{footnote}}
%%\footnotetext{\kern-15.3pt AMS Subject Classification:
%%NNANN; % Symmetric functions
%}
%}}
\maketitle
%%%%%%%%%%%%%%%%%enter abstract, MSCS codes, and keywords here
%%%%%%%%%%%%%%%%%%%%%%%%%%%%%%%%%%%%%%%%%%%%%%%%%%%%%%%%%%%%%%%%%%%%%%%%
\begin{abstract}
The Schr\"odinger-Robertson inequality for relativistic position and momentum operators $X^\mu, P_\nu$, $\mu, \nu = 0,1,2,3$, is interpreted in terms of Born reciprocity and `non-commutative' relativistic position-momentum space geometry. For states which saturate the Schr\"odinger-Robertson inequality, a typology of semiclassical limits is pointed out, characterised by the orbit structure within its unitary irreducible representations, of the full invariance group of Born reciprocity, the so-called `quaplectic' group $U(3,1) \otimes_{\sf s} H(3,1)$ (the semi-direct product of the  unitary relativistic dyamical symmetry $U(3,1)$ with the Weyl-Heisenberg group $H(3,1)$). The example of the `scalar' case, namely the relativistic oscillator, and associated multimode squeezed states, is treated in detail. In this case,
it is suggested that the semiclassical limit corresponds to the separate emergence of space-time and matter, in the form of the stress-energy tensor, and the quadrupole tensor, which are in general  reciprocally equivalent.\\
%{\bf Keywords: }
%Keywords
\end{abstract}
%%%%%%%%%%%%%%%%%%%%%%%%%%%%%%%%%%%%%%%%%%%%%%%%%%%%%%%%%%%%%%%%%%%%%
%%%%%% end title page?
\pagebreak

\section{Introduction and Main Results}
\label{sec:Intro}
Deep insights into fundamental physics often follow from the logical pursuit of theoretical possibilities, as examples such as the nature of charge (Yang-Mills theories, monopoles) and the nature of the vacuum (QCD instantons, anomalies in gauge theories, Hawking radiation) well show. In this letter we adopt the fundamental view of particles and fields advocated almost 50 years ago by Born\cite{Born1949} and Born and Green  \cite{Born1949a,Green1949} via their  `reciprocity principle'.  A modern rendering of this idea \cite{Low2002,Low2005a} is that the relativistic phase space of $X^\mu$ and $P_\nu$, $\mu, \nu = 0,1,2,3$,  given the Heisenberg commutation relations
\begin{align}
{[} X^\mu, P_\nu {]} = & \, i \hbar {\delta^\mu}_\nu I,
\label{eq:HeisenbergStandard}
\end{align}
constitutes a \emph{bona fide} `non-commutative geometry' \cite{Connes1994,Majid1993,Turaev1989}, which should be studied in its own right as a generalisation of position-momentum space. 
Indeed, a feature of Born reciprocity is precisely the blurring of the fundamental distinction between position and momentum via the splitting of Planck's constant into two factors, 
$\hbar = \textsl{a} \cdot \textsl{b}$ corresponding to new constants of  length and momentum respectively.  By contrast, the formulation of Low \cite{Low2002} posited the existence of a related, new fundamental constant (also labelled as $b$), corresponding to a maximum rate of change of momentum: a \emph{force}.  This is the momentum-time equivalent of the maximum rate of change of position (the \emph{velocity} of light) in standard space-time relativity, and there is an elegant formulation of mechanics \cite{Low2005b} in which the Lorentz group of special relativity, consisting of transformations between inertial observers at constant relative speed up to a maximum speed $c$, is extended to the quaplectic group\footnote{For the formal definition see below and  \cite{Low2002,Low2005a,Low2005b}} of
reciprocal relativity encompassing accelerated motion between observers, with maximum relative rate of change of momentum $b$. From this new fundamental constant, we can \emph{define} Newton's gravitational constant \textit{G} in the following way:
\begin{equation} \label{eq:NewtonConstant}
G_N=\alpha _G c^4 /b
\end{equation}
$\alpha_G$ being a scaling constant.  As is well known, purely at the level of dimensional analysis, inclusion of Newton's constant $G_N$ along with $c$ and $\hbar$ allows the construction of universal constants of \textit{any} desired dimension - the `Planck units' of mass, length, time, energy, power, acceleration, etc.  Alternatively, we could derive these scales in terms of $b$, $c$ and $\hbar$. For example, scales for time, position, momentum, energy and acceleration are defined \cite{Low2002}  by
\begin{equation}
\label{eq:PlanckUnits}
\lambda_t=\sqrt{\hbar/bc}, \,  \lambda_x=\sqrt{\hbar c/b}, \, \lambda_p=\sqrt{\hbar b/c}, \, \lambda_e=\sqrt{\hbar bc}, \, \lambda_a = c \sqrt{bc/\hbar }
\end{equation}
up to overall scale factors. If these equal 1, and if $\alpha_G = 1$ in (\ref{eq:NewtonConstant}), then these correspond exactly with the usual Planck quantities.

Using these units, for example, the Heisenberg uncertainty relation takes the symmetrical and dimensionless form
\begin{equation*}
\frac{\Delta X}{\lambda_x}.\frac{\Delta P}{\lambda_p}\geq \frac{1}{2}.
\end{equation*}
This equation is invariant under the `reciprocity' transformation exchanging position and momentum,
\[
\frac{X}{\lambda_x}.\rightarrow \frac{P}{\lambda_p}, \quad \frac{P}{\lambda_p}\rightarrow -\frac{X}{\lambda_x}
\]
and it was the ubiquitous appearance of this (discrete) symmetry (as, for instance, in the Heisenberg relations themselves, and in ordinary nonrelativistic Hamilton equations of motion), which originally led Born to his reciprocity principle \cite{Born1949}.  In fact the interested was in meson masses, and the use of `reciprocal invariant' combinations of $X^\mu$ and $P_\nu$ was advocated as the seed for relativistic wave equations.  For example in the scalar case, the master equation \cite{Born1949}
\begin{equation} \label{eq:BornGreenScalar2}
\left(\frac{X^\mu X_\mu}{{\lambda_x}^2} + \frac{P^\mu P_\mu}{{\lambda_p}^2}\right)\Phi = s\Phi
\end{equation}
(interpreted as a differential equation in momentum space for suitable functions $\Phi(p^0,p^1,p^2,p^3)$ with $P^\mu \rightarrow p^\mu , \ X_\mu \rightarrow + i\hbar \partial / \partial p^\mu$ ) has, as appropriate solutions, eigenfunctions which are generalised Laguerre polynomials in $p^2 \equiv p^\mu p_\mu$ for a discrete eigenvalue spectrum of $s$.  The zeroes of these functions then in principle provide a particle mass spectrum, as an obvious generalisation of the mass shell condition $p^2 = \mu^2  \textsl{p}^2 \equiv (mc)^2$ for a single particle (a spinor version of the master equation was also studied for fermions).

A modern perspective has been developed by Low \cite{Low2002, Low2005a} who has  argued that the theory should be developed along kinematical lines, starting from the appropriate extended relativistic reciprocal invariance group. As discussed in detail in \cite{Low2002} (see also \cite{Low2005b}), this is the so-called quaplectic group $U(3,1) \otimes_{\sf s} H(3,1)$ (the semi-direct product of the  unitary relativistic dyamical symmetry $U(3,1)$ with the Weyl-Heisenberg group $H(3,1)$). In this programme there is a well-defined route to reciprocally-invariant wave equations, and hence the physics of Born reciprocity, via induced representation theory, in direct analogy with the classical development starting firstly with Wigner's analysis of induced 
unitary  irreducible representations of the Poincar\'e group, and subsequently their realisation for example as solutions of Bargmann-Wigner equations (see for example \cite{BekaertBoulanger2003}). Of course, the cases in which \emph{interactions} can be consistently introduced are precisely those needed to date to describe standard physics, namely spin-$\frac 12$, 0 and 1 -- quarks, leptons and gauge and scalar bosons, together with possible supersymmetric partners (like the gravitino and the graviton in the non-renormalisable case of (super)-gravity).

Rather than follow this route directly, in this letter we take up issues of physical states, and the generalised quantum $\rightarrow$ classical limit, assuming that this is meaningful in the context of  reciprocity (notwithstanding the well-known and severe difficulties of handling particle localisation, and defining coordinate operators, in standard field theory).   \S \ref{sec:Quaplectic} below outlines the identification of the quaplectic group $U(3,1) \otimes_{\sf s} H(3,1)$ as the appropriate generalisation of relativistic symmetry to encompass the reciprocity principle. Also, rather than  work with generic quaplectic group unitary irreducible representations, we examine just one basic representation, the `scalar' system provided by the relativistic oscillator itself.  \S \ref{sec:Oscillators} provides details of the oscillator construction, and for this case also introduces various generalised uncertainty relations and  reviews a relativistic version of the well-known multimode squeezed states which saturate them. Specifically, we note that one particular measure of classicality, the degree of saturation of the Schr\"odinger-Robertson inequality, is in fact quaplectic invariant. 
In \S \ref{sec:Discussion} the implications of this observation are explored for general states and specifically, for the
(relativistic) squeezed states which saturate the inequality. It is argued that, within the oscillator representation and potentially in general, there is a `geometry' of semiclassical limits associated with the quaplectic group orbit structure of such states.
\section{The quaplectic group}
\label{sec:Quaplectic}
According to Low \cite{Low2002,Low2005a,Low2005b}, the true (local) dynamical symmetry group of nature should implement the Born reciprocity principle of the interchangeability of position and momentum observables, while subsuming the Heisenberg commutation relations and also special relativity. This group has been identified as the so-called `quaplectic' group $U(3,1)\otimes_{\sf s} H(3,1)$. For present purposes this can be seen as follows.
A defining feature is that there is a quadratic Casimir invariant containing the terms
\begin{equation}
\label{eq:QuadraticCasimirDots}
 \frac{X^\mu X_\mu}{{\lambda_x}^2} + \frac{P^\mu P_\mu}{{\lambda_p}^2} + \cdots
\end{equation}
where $\lambda_x$ and $\lambda_p$ are the position and momentum scales referred to in (\ref{eq:PlanckUnits}), and the algebra includes the canonical commutation relations (compare (\ref{eq:HeisenbergStandard})), 
\begin{equation}
\label{eq:HeisenbergScaled}
\left[\frac{X^\mu}{\lambda_x},\frac{P_\nu}{\lambda_p}\right] = i\alpha_\hbar{\delta^\mu}_\nu I,
\end{equation}
where we write the
central term in the Weyl-Heisenberg algebra as $\alpha_\hbar I$ for some dimensionless scaling constant from (\ref{eq:PlanckUnits}), and  $\hbar=\alpha_\hbar \lambda_x \lambda_p$ (compare this to Born's splitting of $\hbar =  \textsl{a} \cdot \textsl{b}$). 
From equation
(\ref{eq:QuadraticCasimirDots}) it can be seen that transformations amongst $X^\mu$ and $P_\nu$ preserving the quadratic form appearing in $C_2$ must in fact belong to the full 8-dimensional orthogonal group $O(6,2)$. It is further required that they be automorphisms of the Weyl-Heisenberg algebra (\ref{eq:HeisenbergScaled}), namely belong to the canonical group\footnote{
There exist automorphisms (permutations amongst $X^\mu$ and $P_\nu$) which change the signature of the quadratic form on the right-hand side of the commutation relations (\ref{eq:HeisenbergScaled}). Thus the Weyl-Heisenberg algebra $H(3,1)$ is isomorphic to $H(4)$, and the automorphism group is simply $Sp(8,{\mathbb R})\otimes_{\sf s} H(4) \simeq Sp(8,{\mathbb R})\otimes_{\sf s} H(3,1)$ (see also \cite{Folland1989}).}
$Sp(8,{\mathbb R})\otimes_{\sf s} H(4)\approx Sp(8,{\mathbb R})\otimes_{\sf s} H(3,1)$ (the omitted terms  $\cdots$ in
(\ref{eq:QuadraticCasimirDots}) are required to secure invariance also
with respect to the Weyl-Heisenberg group; see (\ref{eq:QuadraticCasimirAll})). Noting 
\begin{align}
Sp(8,{\mathbb R}) \cap & \, O(6,2) \simeq U(3,1),
\end{align}
we have the quaplectic group  $U(3,1)\otimes_{\sf s} H(3,1)$ as the relativistic canonical group of  reciprocity, as claimed: the unique group which contains both the Weyl-Heisenberg group, as well as the Poincar\'{e} group, together with additional transformations between $X^\mu$ and $P^\nu$ with the required properties.

In full, the generators of the quaplectic group are the operators ${E^\mu}_\nu$, $\mu, \nu = 0,1,2,3$, with the hermiticity conditions in unitary representations 
\begin{align}
\label{eq:HermiticityU31}
({E^\mu}_\nu)^\dagger = & \,  \eta^{\mu \rho} \eta_{\nu \sigma} ({E^\sigma}_\rho)
\end{align}
which generate the real Lie algebra of $U(3,1)$, 
\begin{align}
\label{eq:CommutatorsU31}
{[} {E^\mu}_\nu, {E^\rho}_\sigma {]} = & \, ( {\delta_\nu}^\rho {E^\mu}_\sigma -  {\delta^\mu}_\sigma {E^\rho}_\nu),
\end{align}
together with the complex vector operator $Z^\mu$ and its conjugate ${\overline{Z}}\mbox{}^\mu$,
\begin{align}
\label{eq:Hermiticity Z}
(Z^\mu)^\dagger = & \, \eta^{\mu \rho} {\overline{Z}}\mbox{}_\rho \equiv {\overline{Z}}\mbox{}^\mu
\end{align}
which fulfil the Heisenberg algebra (with central generator $\alpha_\hbar I$)
\begin{align}
\label{eq:HeisenbergZ}
{[} Z^\mu, {\overline{Z}}\mbox{}^\nu {]} = & \, - \alpha_\hbar  \eta^{\mu \nu} I,  \nonumber \\
{[} Z^\mu, Z^\nu {]} = &\, 0 \, = {[} {\overline{Z}}\mbox{}^\mu, {\overline{Z}}\mbox{}^\nu {]} 
\end{align}
together with
\begin{align}
\label{eq:CommutatorsEZ}
{[} {E^\mu}_\nu, {\overline{Z}}\mbox{}^\rho {]} = & \, {\delta_\nu}^\rho {\overline{Z}}\mbox{}^\mu, \nonumber \\
{[} {E^\mu}_\nu,  Z^\rho {]} = & \, -  \eta^{\mu \rho}  Z_\nu.
\end{align}
In the above, the Lorentz metric $\eta_{4 \times 4} = diag(+1,-1,-1,-1)$ is adopted together with standard conventions for raising and lowering indices.

Relativistic position and momentum operators $X^\mu$, $P^\nu$ are defined as the quadrature components of $Z^\mu$ and ${\overline{Z}}\mbox{}^\mu$, namely
\begin{align}
\label{eq:Quadratures}
Z^\mu =  \frac{1}{\sqrt{2} }(\frac{X^\mu}{\lambda_x} - i \frac{P^\mu}{\lambda_p}), & \, \quad
\overline{Z}^\mu =  \frac{1}{\sqrt{2} }(\frac{X^\mu}{\lambda_x} +i \frac{P^\mu}{\lambda_p}), \nonumber \\
\frac{X^\mu}{\lambda_x} =  \frac{1}{\sqrt{2} }(Z^\mu + {\overline{Z}}\mbox{}^\mu), &\, \quad 
\frac{P^\mu}{\lambda_p} =  \frac{i}{\sqrt{2} }(Z^\mu - {\overline{Z}}\mbox{}^\mu), \nonumber \\                                                                                
\mbox{with} \qquad {[} \frac{X^\mu}{\lambda_x}, \frac{P_\nu}{\lambda_p}{]} =   
i \alpha_\hbar {\delta^\mu}_\nu,  \quad \mbox{and} & \, \quad 
{[} X^\mu, X^\nu {]} = 0 = {[}  P^\mu, P^\nu {]}
\end{align}
and $\hbar = \alpha_\hbar \lambda_x \lambda_p$ as before.

The structure of the quaplectic algebra is made more transparent in terms of auxiliary generators ${\{} {e^\mu}_\nu {\}}$ which provide a `spin-orbit' like decomposition,
\begin{align}
\label{eq:SpinOrbit}
{e^\mu}_\nu :=  {E^\mu}_\nu - \frac{1}{2 \alpha_\hbar}{\{}{\overline{Z}}\mbox{}^\mu , Z_\nu{\}}, \quad
& \, 
{E^\mu}_\nu =   {e^\mu}_\nu +  \frac{1}{2 \alpha_\hbar}{\{}{\overline{Z}}\mbox{}^\mu , Z_\nu {\}},
\end{align} 
such that the ${e^\mu}_\nu$ satisfy the $U(3,1)$ algebra, but commute with $H(3,1)$:
\begin{align}
\label{eq:EeCommutators}
{[} {E^\mu}_\nu, {e^\rho}_\sigma {]} = & \,  ({\delta_\nu }^\rho{e^\mu}_\sigma - {\delta^\mu}_\sigma {e^\rho}_\nu),  \nonumber \\
{[} {e^\mu}_\nu, {e^\rho}_\sigma {]} = & \,  ({\delta_\nu }^\rho{e^\mu}_\sigma - {\delta^\mu}_\sigma {e^\rho}_\nu), \nonumber \\
  {[} {e^\mu}_\nu, Z^\rho {]} = &\, 0 =  {[} {e^\mu}_\nu, \overline{Z}^\rho {]}.
\end{align}
Subject to technicalities of Mackey induced representation theory, (see \cite{Low2005a}), it is clear that a generic unitary irreducible representation (unirrep) of the quaplectic group can be associated with the tensor product of a unirrep of $U(3,1)$ (provided by nonzero ${e^\mu}_\nu$), with suitable unirrep(s) of the Weyl-Heisenberg algebra $H(3,1)$. The latter can of course themselves be identified via induced representations \cite{Wolf1975}.

The quaplectic algebra itself can be re-written in a tensor form which identifies its Lorentz (and Poincar\'{e}) subalgebras. Defining $E_{\mu \nu} \equiv \eta_{\mu \rho}{E^\rho}_\nu$ and ${F^\epsilon}_{\mu\nu} = E_{\mu \nu}  - \epsilon E_{\nu \mu}$, $\epsilon = \pm 1$,  we have 
\[ 
{[}{F^\epsilon}_{\mu\nu} , {F^\epsilon{}'}_{\rho \sigma}{]} =
\left( \eta_{\nu \rho}F^{\epsilon \epsilon{}'}_{\mu \sigma} - \epsilon \eta_{\nu \rho}F^{\epsilon \epsilon{}'}_{\mu \sigma}
-  \epsilon{}' \eta_{\nu \sigma}F^{\epsilon \epsilon{}'}_{\mu \rho}+  \epsilon\epsilon{}' \eta_{\mu \sigma}F^{\epsilon \epsilon{}'}_{\nu \rho} \right),
\]
or specifically with $L_{\mu \nu} = i( E_{\mu \nu} - E_{\nu \mu})$, $M_{\mu \nu} = E_{\mu \nu} + E_{\nu \mu}$,
\begin{align}
\label{eq:ExtendedQplecticCR}
{[} L_{\kappa \lambda}, L_{\mu \nu} {]} = & \, i\left( \eta_{\lambda \mu}  L_{\kappa \nu} - \eta_{\kappa \mu} L_{\lambda \mu}  - \eta_{\lambda \nu}  L_{\kappa \mu} + \eta_{\kappa \nu}  L_{\lambda \mu} \right),  \nonumber \\
{[} L_{\kappa \lambda}, M_{\mu \nu} {]} = & \,i\left( \eta_{\lambda \mu}  M_{\kappa \nu} - \eta_{\kappa \mu} M_{\lambda \mu}  + \eta_{\lambda \nu}  M_{\kappa \mu} - \eta_{\kappa \nu}  M_{\lambda \mu} \right),  \nonumber \\
{[} M_{\kappa \lambda}, M_{\mu \nu} {]} = & \, \left( \eta_{\lambda \mu}  M_{\kappa \nu} + \eta_{\kappa \mu} M_{\lambda \mu}  + \eta_{\lambda \nu}  M_{\kappa \mu} + \eta_{\kappa \nu}  M_{\lambda \mu} \right),  \nonumber \\
{[} L_{\kappa \lambda}, X_\mu{]} = & \, i\left(\eta_{\lambda \mu} X_\kappa - \eta_{\kappa \mu} X_\lambda\right) , \nonumber \\
{[} L_{\kappa \lambda}, P_\mu{]} = & \, i\left(\eta_{\lambda \mu} P_\kappa - \eta_{\kappa \mu} P_\lambda\right) , \nonumber \\
{[} M_{\kappa \lambda}, X_\mu{]} = & \,- i\left(\eta_{\lambda \mu} P_\kappa - \eta_{\kappa \mu} P_\lambda\right) , \nonumber \\
{[} M_{\kappa \lambda}, P_\mu{]} = & \, i\left(\eta_{\lambda \mu} X_\kappa - \eta_{\kappa \mu} X_\lambda\right).
\end{align}
Beyond the Lorentz algebra generated by the $L_{\mu \nu}$, identification of  the Poincar\'{e} subalgebra obviously depends on the choice of abelian four-vector operator,
for example $X_\mu$ or  $P_\mu$.

Henceforth we restrict ourselves to the `scalar' case with 
\begin{align}
\label{eq:ScalarDefn}
{e^\mu}_\nu \equiv 0, & \qquad {E^\mu}_\nu = \frac{1}{2\alpha_\hbar}{\{}{\overline{Z}}\mbox{}^\mu, Z_\nu {\}} .
\end{align}
(more generally we could add and subract from either matrix real constant times the identity matrix, $\frac 14 \varepsilon {\delta^\mu}_\nu I$). In physical terms the appropriate unitary representation of the Weyl-Heisenberg algebra $H(3,1)$ is associated with the so-called  `relativistic oscillator' \cite{Dirac1945} where generators are given via actions on suitable functions on coordinate space (or on holomorphic functions on a complex normed space, so-called Bargmann space). We have from (\ref{eq:Quadratures})
\begin{align}
\label{eq:CreationAnnihilationDefns}
Z_0 = \frac{1}{\sqrt{2}}(\frac{X_0}{\lambda_x} - i \frac{P_0}{\lambda_p }), \quad & \, Z_i = \frac{1}{\sqrt{2}}(\frac{X_i}{\lambda_x} - i \frac{P_i}{\lambda_p }).
\end{align}
In terms of standard coordinates ($P_\mu \rightarrow -i \hbar \partial /\partial x^\mu$), $Z_0$ is identified with a `creation' operator ${a_0}^\dagger$, whereas $Z_i$ is identified with an `annihilation' operator $a_i$, $i=1,2,3$ (reflecting the sign change in the commutation relations of $Z^\mu$ and ${\overline{Z}}\mbox{}^\nu$ between the temporal and spatial parts; note that $x_0 = x^0, x_i = -x^i$). The zero occupancy state in the number basis is for example\cite{Dirac1945,Kim1991}
\begin{align}
\label{eq:ZeroOccupancy}
\Psi_{n_0\!=\! n_1\!=\! n_2\!=\! n_3\!=\!0}(x^\mu)  \propto & \,  \exp{(\!-\!\textstyle{\frac 12} (x^0)^2)} \cdot 
\exp{(\!+\!\textstyle{\frac 12} \sum_{i=1}^3 x^i x^j \eta_{ij})} \nonumber \\
= & \, \exp{ \left( -{\textstyle{\frac 12}}((x^0)^2\!+\!(x^1)^2\!+\!(x^2)^2\!+\!(x^3)^2)\right) }.
\end{align}

Finally, in relation to the general quaplectic algebra note that the spin-orbit decomposition (\ref{eq:SpinOrbit}) allows for an easy identification  of  Casimir operators of Gel'fand type \cite{Low2002}. Define
\begin{align}
\label{eq:CasimirGeneric}
{(e^{(n+1)})^\mu}_\nu = & \, {(e^{(n)})^\mu}_\rho  {e^\rho}_\nu , \quad {(e^{( 1)})^\mu}_\nu \equiv {e^\mu}_\nu, \nonumber \\
\mbox{then} \quad  C_{n} =  & \, \mbox{tr}(e^{(n)}) =  {(e^{(n)})^\mu}_\mu;
\end{align} 
tensorially (from (\ref{eq:EeCommutators}a) these traces are $U(3,1)$ invariants; however from 
(\ref{eq:EeCommutators}c) they are trivially also quaplectic Casimirs. Explicitly, we have for example
(compare (\ref{eq:QuadraticCasimirDots}))
\begin{align}
\label{eq:QuadraticCasimirAll}
 - C_1 = & \, \frac 12 \left( \frac{X^\mu X_\mu}{\lambda_x^2} + \frac{P^\mu P_\mu}{\lambda_p^2}\right)   - \alpha_\hbar {\mathcal N} - 4 {\alpha_\hbar}^2 I,
\end{align}
where ${\mathcal N} = {E^\mu}_\mu$; the remaining independent Casimirs $C_2$ and $C_3$ can similarly be evaluated. For the scalar representation,  the Casimirs $C_1$, $C_2$, $C_3$ are identically zero, or more generally take constant values $\varepsilon$, $\frac 14 \varepsilon^2$,
$\frac {1}{16} \varepsilon^3$, $\cdots$ respectively on all states.  

\section{Relativistic oscillators, squeezed states and the Schr\"odinger-Robertson inequality}
\label{sec:Oscillators}
It is well known that the Heisenberg uncertainty relation is but one of a hierarchy of inequalities relating means and variances of observables. Stronger than the Heisenberg relation $\Delta X \Delta P \ge \frac 12 \hbar$ for canonical operators (but equivalent for uncorrelated states) is the Schr\"odinger \cite{Schroedinger1930} inequality for two observables A, B, 
\begin{align}
\Delta^2 A \cdot \Delta^2 B \ge & \, \frac 14 | \langle   {[} A, B {]} \rangle |^2 + \mbox{cov}(A,B), \nonumber \\
\mbox{where} \quad \mbox{cov}(A,B) = & \, \frac 12 ( \langle A B \rangle + \langle B A \rangle )- \langle A  \rangle \langle B \rangle 
\quad \mbox{and} \quad \Delta^2(A) = \mbox{cov} (A,A).
\end{align}
The latter itself is a special case of the more general Schr\"odinger-Robertson inequality 
\cite{Robertson1934} for $2n$ observables $A_1, A_2, \cdots, A_{2n}$,
\begin{align}
\label{eq:SchroedRob}
\det(\Sigma) \ge  \det (C), \quad \mbox{where} \quad & \, \Sigma_{kl} = \mbox{cov}(A_k, A_l), \quad C_{kl} = - \frac i2 \langle {[} A_k, A_l {]} \rangle.
\end{align}                                                                                     

For the canonical observables associated with a relativistic oscillator in 4 dimensions, namely $X^0$, $X^1$, $X^2$, $X^3$ and   $P^0$, $P^1$, $P^2$, $P^3$, the same considerations apply. Using the extended index notation $\mu, \nu$  $= 0,1,2,3$, and $\mu', \nu' = \mu + 4, \nu+4$  $ = 4,5,6,7$, define the $8 \times 8$ matrices
\begin{align}
\label{eq:8times8defns}
\Sigma = \left( \begin{array}{cc} \Sigma^{\mu \nu} & \Sigma^{\mu \nu'} \\ \Sigma^{\mu' \nu} & \Sigma^{\mu' \nu'} \end{array} \right)
 = & \,  \left( \begin{array}{cc} \mbox{cov}(X^\mu, X^\nu) & \mbox{cov}(X^\mu, P^\nu) \\ \mbox{cov}(P^\mu, X^\nu) & \mbox{cov}(P^{\mu}, P^{\nu})\end{array} \right), \nonumber \\
C = \left( \begin{array}{cc} C^{\mu \nu} & C^{\mu \nu'} \\ C^{\mu' \nu} & C^{\mu' \nu'} \end{array} \right) 
= & \, \left( \begin{array}{cc}-\frac{i}{2}\langle{[}X^{\mu}, X^{ \nu} {]}\rangle & -\frac{i}{2}\langle{[}X^{\mu}, P^{ \nu} {]}\rangle  \\ -\frac{i}{2}\langle{[}P^{\mu}, X^{ \nu} {]}\rangle & -\frac{i}{2}\langle{[}P^{\mu}, P^{ \nu} {]}\rangle\end{array} \right) 
= { \textstyle \frac 12} \hbar \left( \begin{array}{cc} 0 &  \eta^{\mu \nu} \\ -\eta^{\mu \nu} & 0\end{array} \right);
\end{align}
then according to (\ref{eq:SchroedRob}) above (as $\det(C) = (\frac 12 \hbar)^8$) we have in any state $|\psi \rangle$ 
\begin{align}
\label{eq:OscillatorSchroedRob}
\det(\Sigma) \ge & \, ( { \textstyle \frac 12}  \hbar)^8.
\end{align}

From the point of view of recicprocity, it is important to characterise the dependence of $\det (\Sigma)$ on the quaplectic group. For an infinitesimal transformation belonging to $U(3,1)$ corresponding to $|\psi \rangle \rightarrow U|\psi \rangle$, 
where
\begin{align}
U = e^{({\omega^\mu}_\nu {E^\nu}_\mu)} \simeq I + {\omega^\mu}_\nu {E^\nu}_\mu, \quad &\, 
U^\dagger = U^{-1} = e^{-({\omega^\mu}_\nu {E^\nu}_\mu)} \simeq I - {\omega^\mu}_\nu {E^\nu}_\mu,
\end{align}
we have (by a slight abuse of notation)
\begin{align}
\label{eq:SigmaVariation}
\det (\Sigma) \rightarrow \det (U^{-1}\Sigma U)= \det(\Sigma - {[}{\omega^\mu}_\nu {E^\nu}_\mu, \Sigma{]}).
\end{align}
It is shown explicitly in the appendix, \S A, that $\det (\Sigma)$ is $U(3,1)$ invariant, and given that a similar calculation for the Weyl-Heisenberg group goes through, is in fact invariant under the full quaplectic group\footnote
{
It is pointed out that in fact there are a number of $U(3,1)$ invariants associated with expected values of polynomials in the generators, including a subset of quaplectic invariant ones, all of which could potentially give interesting
information on semiclassical limits. See  \S A for details of these more general invariants.}.
\section{Discussion}
\label{sec:Discussion}
The quaplectic invariance of $\det(\Sigma)$ can be related to the structure of relativistic position-momentum space as follows. Firstly, for uncorrelated states with vanishing $\mbox{cov}(X^\mu, X^\nu)$, $\mbox{cov}(P^\mu, P^\nu) $ and  $\mbox{cov}(X^\mu, P^\nu)$ if $\mu \ne \nu$, we have (taking the square root of (\ref{eq:OscillatorSchroedRob}))\footnote{For an even number of observables $\det(C)$ is of course the square of a pfaffian.},
\begin{align}
\Delta X^0 \Delta X^1 \Delta X^2 \Delta X^3 \Delta P^0 \Delta P^1 \Delta P^2 \Delta P^3 \ge ( { \textstyle \frac 12}  \hbar)^4
\end{align}
-- a natural extension of Heisenberg's principle to the case of `space-time granularity' (see also \cite{Robertson1934,Synge1971}). 
However, the invariance of $\det(\Sigma)$ shows that there are many `physically equivalent' states $|\psi'\rangle = U|\psi \rangle$,
related by quaplectic unitary transformations, with the same measure of spreading. As these certainly include correlated states, including those with position-momentum correlations, the claim of Born reciprocity to set position and momentum `coordinates' on an equal footing, is once again demonstrated -- for a given uncertainty measure $\sigma$, there is no natural distinction between uncorrelated and correlated states.

Let us consider the $8 \times 8$ covariance matrix $\Sigma$ in more detail (compare \cite{Trifonov1997}). It is easily established using elementary algebra that $\Sigma$ is a positive semi-definite matrix, $\Sigma \ge 0$. Moreover as $\det(C) > 0$, we have that $\det(\Sigma) >0$ and so, lacking zero eigenvalues, $\Sigma$ is in fact positive definite, $\Sigma > 0$. Under these circumstances a well-known result in matrix algebra \cite{Gantmacher1959} states that $\Sigma$ can be diagonalised by an $8 \times 8$ symplectic matrix\footnote{
Note that $\Sigma$, being positive semi-definite, can always be diagonalised by an orthogonal transformation, but we do not need this property for the present discusson. In (\ref{eq:SigmaDiagonalisation}) the symplectic metric $J$ has been chosen to show the role played by the Lorentz metric $\eta_{4 \times 4} = diag(+1,-1,-1,-1)$, although $J$ is of course equivalent to a standard antisymmetric matrix with $1_{4 \times 4}$ replacing $\eta_{4 \times 4}$ by an elementary row and column permutation.} $S$,
\begin{align}
\label{eq:SigmaDiagonalisation}
\Sigma' = S \Sigma S^T \quad & \, \mbox{with} \quad S J  S^T = J, \quad J =
\left( \begin{array}{cc} 0 & \eta_{4 \times 4} \\ -\eta_{4 \times 4} & 0 \end{array} \right).
\end{align}
Such a symplectic transformation $S$ on the matrix $\Sigma$ can also be implemented by an appropriate \emph{state} transformation on $|\psi\rangle$, by a suitable operator belonging to the natural unitary representation of the real symplectic group carried by the oscillator states (there is no similar way available of implementing the orthogonal transformation which diagonalises $\Sigma$). Extending the notation of (\ref{eq:8times8defns}), the symplectic algebra is defined as follows. Introduce the compound index $K, L = 0,1,\ldots,7$, with $\kappa = 0,1,2,3$, $\kappa' = 4,5,6,7$, and so on, and therewith the 8-component vector
\begin{align}
\label{eq:8componentZ}
Z_K = \left( \begin{array}{c} Z_\kappa \\ Z_{\kappa'} \end{array} \right) = & \, \left( \begin{array}{c} Z_\kappa \\ \overline{Z}\mbox{}_{\kappa} \end{array} \right).
\end{align}
Then the generators $Z_K$ together with $Z_{KL} = \frac {1}{2\alpha_\hbar} {\{} Z_K, Z_L {\}}$ fulfil 
\begin{align}
\label{eq:CanonicalGroup}
{[} Z_{KL}, Z_{MN} {]} = & \, -\left(J_{LM}Z_{KN} + J_{KM}Z_{LN}+J_{LN}Z_{KM}+J_{KN}Z_{LM}\right), \nonumber \\
{[} Z_{KL}, Z_{M} {]} = & \, -\left(J_{LM}Z_K + J_{KM}Z_L\right), \nonumber \\
{[} Z_K, Z_L {]} = & \, -\alpha_\hbar J_{KL} I
\end{align}
where the $8 \times 8$ symplectic metric $J$ is defined in (\ref{eq:SigmaDiagonalisation}); note from (\ref{eq:8times8defns}) that $C = - \frac 12 \hbar J$. The commutation relations
(\ref{eq:CanonicalGroup}) define the semidirect product $Sp(8,{\mathbb R}) \otimes_{\sf s} H(3,1)$ of the real 8-dimensional symplectic canonical algebra, with the Weyl-Heisenberg algebra\footnote{
In terms of Lorentz-covariant notation, the generators are thus $Z_{\mu\nu} =  {\frac {1}{2\alpha_\hbar}}{\{}Z_\mu,Z_\nu{\}}$,
${Z}_{\mu' \nu'} =  {\frac {1}{2\alpha_\hbar}}{\{}\overline{Z}_\mu,\overline{Z}_\nu{\}}$, and $Z_{\mu' \nu} \equiv E_{\mu \nu}$ (see 
(\ref{eq:ExtendedQplecticCR})).
}. Thus the diagonalisation of $\Sigma$ is implemented by a unitary symplectic transformation of the form
\begin{align}
|\psi' \rangle = U |\psi \rangle, \quad & \, U = e^{\frac 12 \theta^{KL}Z_{KL}}.
\end{align}

Finally, we consider the states $|\psi\rangle$ for which the Schr\"{o}dinger-Robertson inequality (\ref{eq:SchroedRob}) is saturated (Schr\"{o}dinger-Robertson minimal uncertainty states). It is well known in quantum optics that for systems of several oscillators, these are the so-called multimode squeezed states \cite{Trifonov1997}. The same construction\footnote{In fact in this case of canonical observables, the sqeezed states are equivalent to the well-known Barut-Girardello group-like coherent states (see for example \cite{Trifonov1997}).}
essentially goes through for the present `relativistic oscillator' case (as emphasised above, the respective Weyl-Heisenberg algebras, for standard versus `relativistic' oscillators, are in fact isomorphic, $H(3,1) \simeq H(4)$). The multimode squeezed states are of the form
\begin{align}
\label{eq:MultimodeSqueezed}
|\varphi, \zeta \rangle = & \, e^{\frac 12 \varphi^{KL}Z_{KL}} \cdot e^{\zeta^K Z_K} \cdot |0 \rangle
\end{align}
for some parameters $\varphi^{KL}$ and $\zeta^K$, where the zero occupancy state was defined in 
(\ref{eq:ZeroOccupancy}) above,
\[
 \langle x^0, x^1,x^2,x^3 |0 \rangle = \Psi_{n_0= n_1= n_2= n_3=0}(x^0, x^1,x^2,x^3).
\]
Now in the context of  reciprocity it must be assumed that different, but unitarily equivalent, minimal uncertainty states $|\varphi, \zeta \rangle$, 
$|\varphi', \zeta' \rangle$,   \emph{only correspond to physically distinct semi-classical limits if 
the unitary transformation relating them does} not \emph{belong to the quaplectic group $U(3,1) \otimes_{\sf s} H(3,1)$}. This is the main result of this paper and follows irrevocably from the reciprocity principle and quaplectic symmetry. Thus we conjecture that in the case of a scalar quaplectic system, there are various physically different classes of minimal uncertainty states, parametrised by the geometry of the homogeneous space 
$Sp(8, {\mathbb R})\otimes_{\sf s} H(4)/U(3,1)\otimes_{\sf s} H(3,1) \simeq Sp(8, {\mathbb R})/U(3,1)$.

Insight into the nature of these classes of minimal uncertainty states can be gained by reference to discussions of semicassical limits in ordinary (non-relativistic) quantum mechanics. Recall that general oscillator coherent states, constructed as in (\ref{eq:MultimodeSqueezed}) can also be characterised by their property of diagonalising a linear combination of the annihilation operator and creation operators,
\begin{align}
\label{eq:StandardCoherent}
{\texttt a} a + {\texttt b} a^\dagger |z \rangle = & \,  z |z \rangle,
\end{align}
and for such a system we can of course (to the extent that these are sharp subject to the Heisenberg principle) associate quantities `position' and `momentum' via the expectation values $\langle X \rangle$,
$\langle P \rangle$ (taking the right hand side of (\ref{eq:CreationAnnihilationDefns}) and its conjugate as the definition of $a$, $a^\dagger$ in order to express these in terms of  $z$, $\overline{z}$,  ${\texttt a} $ and ${\texttt b} $ as in (\ref{eq:StandardCoherent})).
`Pure' position and momentum states should in this context be thought of as singular (non-normalisable) limits of such $|z\rangle$, for inadmissible values of the ratio ${\texttt a} /{\texttt b} $\cite{Trifonov1997}. However, from the viewpoint of quaplectic transformations which can rotate unitarily between $X$ and $P$, the attribution of these quantities, regarded as having sharp classical values or not, is purely frame-dependent.

A comparable discussion can be carried through for a plausible interpretation of the `semiclassical' limiting states (\ref{eq:MultimodeSqueezed}), with a view to understand what attributes \emph{can} be regarded as intrinsic, rather than frame dependent.
Firstly note that an obvious set of orbits for $Sp(8, {\mathbb R})/U(3,1)$ is parametrised by the states
\begin{align}
\label{eq:OrbitClasses}
| \Phi \rangle = & \, e^{\frac 12 \varphi^{\mu \nu} Z_{\mu \nu}}\cdot
                                e^{\frac 12 \overline{\varphi}^{\mu \nu} \overline{Z}_{\mu \nu}} \cdot |\Psi \rangle.
\end{align}
for some fixed cyclic vector $|\Psi \rangle$.
However, the appropriate `semiclassical' limit to consider is now $\hbar \rightarrow 0$, governing the transition from quantum to classical regimes, together with $b \rightarrow \infty$, governing the transition from reciprocal relativistic to special relativistic regimes. As expected (see below), the quaplectic group contracts to the Poincar\'{e} algebra
generated by $L^{\circ}_{\mu \nu}$ and $P^{\circ}_\mu$  (or  $X^{\circ}_\mu$) augmented by a set of ten abelian generators $M^{\circ}_{\mu \nu}$ (the contraction limit of the $M_{\mu \nu}$ of 
(\ref{eq:ExtendedQplecticCR})). We infer that in the 
$b \rightarrow \infty$ regime, states of a quaplectically covariant system will be characterised by some eigenvalue $m_{\mu\nu}$ of this tensor. In the scalar case (see (\ref{eq:ExtendedQplecticCR}), (\ref{eq:SpinOrbit}), (\ref{eq:ScalarDefn}), (\ref{eq:Quadratures}), this tensor is essentially $X_\mu X_\nu + P_\mu P_\nu $. A natural further question is then to identify the relevant $b \rightarrow \infty$ contraction limit of the associated symplectic algebra, given the behaviour of its $U(3,1)$ subalgebra. From the commutation relations (from (\ref{eq:CanonicalGroup}))
\begin{align}
\label{eq:ZZbarMCommutator}
{[} \overline{Z}_{\mu \nu} , Z_{\rho \sigma}{]} = & \, \eta_{\mu \rho} M_{\nu \sigma} +
\eta_{\nu \rho} M_{\mu \sigma} +\eta_{\mu \sigma} M_{\nu \rho} +\eta_{\nu \sigma} M_{\mu \rho}, \end{align}
there are two possibilities, involving the decomposition of $M_{\mu\nu}$ (see (\ref{eq:ExtendedQplecticCR})) into its traceless part,
\begin{align}
\label{eq:ContractionTypes}
M_{\mu\nu} = N_{\mu\nu} + \textstyle{\frac 12 }\eta_{\mu\nu}{\mathcal N} ,   \quad 
N_{\mu\nu} :=  M_{\mu\nu} - \textstyle{\frac 12 }\eta_{\mu\nu} {\mathcal N}, & \qquad {\mathcal N} := \frac 12 {M^\nu}_\nu = {E^\nu}_\nu;  \\
\mbox{with} \qquad L_{\mu \nu} =   {L^\circ}_{\mu\nu}; \qquad
{\overline{Z}}_{\mu \nu} =  {\overline{Z}{}^\circ}_{\mu \nu}/b, \quad & \quad  Z_{\mu \nu}= {Z^\circ}_{\mu \nu}/b;  \nonumber \\
\mbox{ and \emph{either}} \qquad \qquad (I) \qquad  M_{\mu \nu}=  {M^\circ}_{\mu\nu} / b, \quad &\mbox{\emph{or}} \quad (I\! I) \qquad 
 {N_{\mu\nu}}=  {{N^\circ}_{\mu\nu}}/b, \, \, {\mathcal N} = {\mathcal N}^\circ / b^2.
 \nonumber
\end{align}                                                                             
In case $(I)$ the ${Z^\circ}_{\mu \nu} , \overline{Z}{}^\circ{}_{\mu \nu} $ become abelian, and 
appropriate physical states could be expected to be eigenstates with some complex eigenvalues $z_{\mu\nu},  \overline{z}_{\mu \nu} $.
In  case $(I\!I)$ these generators fulfil a ten-dimensional Heisenberg algebra with central generator ${\mathcal N}^\circ$ (from (\ref{eq:ZZbarMCommutator}), (\ref{eq:ContractionTypes})) ,
 \begin{align}
 {[} {Z^\circ}_{\mu \nu} , \overline{Z}{}^\circ{}_{\rho \sigma} {]} = & \, 
            ( \eta_{\mu \rho}  \eta_{\nu \sigma} + \eta_{\mu \sigma} \eta_{\nu \rho}) {\mathcal N^\circ}.       \nonumber
\end{align}
Then the states (\ref{eq:OrbitClasses}) (with $|\Psi \rangle = |0\rangle$) have precisely the structure of standard `coherent' states, this time associated with expectation values
of the operators $\langle {Z^\circ}_{\mu \nu}\rangle$ and $\langle \overline{Z}{}^\circ{}_{\mu \nu} \rangle$. 
Using (\ref{eq:CanonicalGroup}) and (\ref{eq:Quadratures}), either case\footnote{
Without the $b \rightarrow \infty$ limit, the states (\ref{eq:OrbitClasses}) can be discussed in the context of group coherent states
(see \cite{Trifonov1997}).}
thus leads to consideration (for scalar systems) of the sharp quantities
$R_{\mu\nu} = \langle X_\mu P_\nu + X_\nu P_\mu \rangle$, and $\langle X_\mu X_\nu - P_\mu P_\nu \rangle$,  together with \emph{eigenvalues} $m_{\mu\nu}$ of $X_\mu X_\nu +P_\mu P_\nu$ in this $b \rightarrow \infty$ regime. Thus, to the state of the quaplectic system can  be associated definite attributes $Q_{\mu \nu} = \langle X_\mu X_\nu \rangle$ and $T_{\mu \nu} = \langle P_\mu P_\nu \rangle $ separately, as well as $R_{\mu\nu}$. It is reasonable to suggest therefore that the significance of the orbit geometry in characterising physically distinct `semiclassical' limits of a quaplectic system is the emergence of structure, in the sense of a quadrupole tensor $Q_{\mu \nu}$ and an energy-momentum tensor $T_{\mu\nu}$ ($R_{\mu\nu} $ perhaps relates to angular momentum). In contrast, the
distinction between  the position and momentum space quantities $\langle X_\mu \rangle$, $\langle P_\nu\rangle$, in the reciprocal-relativistic regime, is frame-dependent ($Z^\mu$ and $\overline{Z}{}^\mu$ transform as four dimensional irreducible representations of the $U(3,1)$ group and acquire additive shifts under the Weyl-Heisenberg group).

In conclusion, we reiterate the claim that the full implications of Born reciprocity are in the direction of a far-reaching generalisation of relativistic physics. The identification of classes of physically distinct quaplectic-invariant minimal uncertainty states in this work sheds light on the geometrical basis of semi-classical limits, and hence on the emergence of a conventional picture of dynamics on space-time, from the generalised reciprocal-invariant  setting, in which position and momentum are interchangeable and their values frame-dependent. 
As a final remark, it should be emphasised that, although the originators did not necessarily have this perspective, it is natural to surmise that the Born reciprocity, as a generalisation of relativistic quantum theory, may also open new avenues towards quantum gravity. Indeed, the natural role \cite{Low2005b} of a unit of \emph{force} (as opposed to simply length or momentum, or acceleration) as a new universal constant of nature may be taken as a prescient hint of an eventual accommodation with the equivalence principle.

\paragraph{Acknowledgement:} Part of this work was done under an Erskine Fellowship appointment held by PDJ  at the University of Canterbury, New Zealand. PDJ would like to thank the Department of Physics and Astronomy, University of Canterbury,  for hospitality, and also the orthopedic surgeons and nurses of Timaru and Christchurch Public Hospitals, for excellent treatment and care, during the preparation of this paper. PDJ acknowledges the Australian Research Council, project DP0208808, for partial support. SOM acknowledges the support of an Australian Postgraduate Award. PDJ and SOM thank Dr Stephen Low for  generously making available to us his working notes on reciprocity, and for discussions on the present work.
%\vfill
%\pagebreak

\begin{appendix}
\section{Appendix: Quaplectic invariance of covariance matrix $\sigma = \det(\Sigma)$ }
\def\theequation{\Alph{section}-\arabic{equation}}
As discussed in \S \ref{sec:Oscillators} above we are interested in the variation of expectation values such as $\mbox{cov}(A,B)$ under quaplectic state transformations 
$|\psi \rangle \rightarrow U|\psi \rangle$.  Firstly for elements of $U(3,1)$ of the form 
\begin{align}
U = e^{({\omega^\mu}_\nu {E^\nu}_\mu)} \cong I + {\omega^\mu}_\nu {E^\nu}_\mu, \quad &\, 
U^\dagger = U^{-1} = e^{-({\omega^\mu}_\nu {E^\nu}_\mu)} \simeq I - {\omega^\mu}_\nu {E^\nu}_\mu,
\end{align}
and therefore $U^{-1} A U \simeq A + \delta A$ with $\delta A = -{[} {\mathcal E}, A {]}$ and 
${\mathcal E}\equiv {\omega^\mu}_\nu {E^\nu}_\mu$. Similarly
we have from  (\ref{eq:SigmaVariation}) (by a slight abuse of notation)
\begin{align}
\label{eq:SigmaVariation2}
\det (\Sigma) \rightarrow \det (U^{-1}\Sigma U)\simeq \det(\Sigma + \delta \Sigma) =  \det(\Sigma - {[}{\mathcal E}, \Sigma{]}).
\end{align}
Moreover,
\begin{align}
\label{eq:DeltaCov}
\delta(\mbox{cov}(A,B)) = & \, \mbox{cov}( \delta A, B) + \mbox{cov}( A, \delta B). 
\end{align}
Evaluating these brackets for the components of $\Sigma_{\mu \nu}$ separately gives
from (\ref{eq:CommutatorsEZ}) and (\ref{eq:Quadratures})
${[} {\mathcal E}, X^\mu {]} =  - i {\omega^\mu}_\sigma P^\sigma$, and 
${[} {\mathcal E}, P^\mu {]} =   + i {\omega^\mu}_\sigma  X^\sigma$.
Thus from (\ref{eq:8times8defns}) we have directly 
\begin{align}
\label{eq:DeltaSigmaExplicit}
\delta \Sigma = & \,  \left( \begin{array}{cc} \delta \Sigma^{\mu\nu} &  \delta \Sigma^{\mu \nu'} \\
                                        \delta \Sigma^{\mu' \nu}  & \delta \Sigma^{\mu' \nu'} \end{array} \right) \nonumber \\
=  & \,  \left( \begin{array}{cc}  i{\omega^\mu}_\sigma \mbox{cov}(P^\sigma, X^\nu) + i{\omega^\nu}_{\sigma'} \mbox{cov}(X^\mu,P^\sigma) &     
                                   i{\omega^\mu}_\sigma \mbox{cov}(P^\sigma, P^{\nu'}) - i{\omega^{\nu'}}_{\sigma'} \mbox{cov}(X^\mu,X^\sigma)  \\
                                 - i{\omega^{\mu '}}_\sigma \mbox{cov}(X^\sigma, X^{\nu} ) + i{\omega^\nu}_{\sigma'} \mbox{cov}(P^{\mu'},P^\sigma)  &   
         - i{\omega^{\mu'}}_\sigma \mbox{cov}(X^\sigma, P^{\nu'} ) - i{\omega^{\nu'}}_{\sigma} \mbox{cov}(P^{\mu'},X^\sigma)  \end{array} \right)                                         \nonumber \\
                        =   & \,   \left(  \begin{array}{cc}    0 & i{\omega^\mu}_{\sigma'} \\
                                         -i {\omega^{\mu'}}_\sigma   &   0    \end{array} \right)  \cdot  
                                    \left( \begin{array}{cc}  \Sigma^{\sigma \nu} &  \Sigma^{\sigma \nu'} \\
                                        \Sigma^{\sigma'  \nu} & \Sigma^{\sigma'  \nu'} \end{array} \right)   +
                          \left( \begin{array}{cc}  \Sigma^{\mu  \sigma }  &  \Sigma^{\mu \sigma'}  \\
                                        \Sigma^{\mu' {} \sigma} &  \Sigma^{\mu' \sigma' } \end{array} \right)  \cdot
                           \left(  \begin{array}{cc}      0 & - i{\omega^\nu}_\sigma \\
                                         +i {\omega^{\nu'}}_{\sigma' }  &    0    \end{array} \right).                  
\end{align} 
Thus 
\begin{align}
\label{eq:detSigmaSchematic}
\Sigma + \delta \Sigma \cong & \, \Sigma + \Omega \cdot \Sigma + \Sigma \cdot \widetilde{\Omega}; \nonumber \\
\det(\Sigma + \delta \Sigma) =  \det(\Sigma) \det(1 + \Sigma^{-1}\delta \Sigma)
           = & \, \det(\Sigma) \cdot \det( 1 + \Sigma^{-1} \Omega \Sigma + \widetilde{\Omega}) \nonumber \\
          = & \, \det(\Sigma) \cdot \exp \mbox{tr} \log (1 + \Sigma^{-1} \Omega \Sigma + \widetilde{\Omega}) \nonumber \\
          \cong & \,  \det(\Sigma) \cdot \exp \mbox{tr} (\Sigma^{-1} \Omega \Sigma + \widetilde{\Omega}) \nonumber \\
          = & \, \det(\Sigma) \cdot \exp \mbox{tr} ( \Omega  + \widetilde{\Omega})  \nonumber \\
          \equiv & \, \det(\Sigma)  
\end{align}
since both $\Omega$ and $\widetilde{\Omega}$ are manifestly traceless. 

For the corresponding calculation for elements of the Weyl-Heisenberg group,   
\begin{align}
U = & \, e^{(\zeta^\mu Z_\mu- \overline{\zeta}^\mu \overline{Z}_\mu)} \cong I + 
                {\zeta^\mu}Z_\mu - \overline{\zeta}^\mu \overline{Z}_\mu,   
\end{align}                      
in the evaluation of terms of the form (\ref{eq:DeltaCov}), 
the contributions to $\delta A, \delta B$ are $\propto I$ (because of the Heisenberg commutation relations).
In consequence, such terms give identically vanishing covariances. Thus the invariance of $\det(\Sigma)$ 
under the whole quaplectic group\footnote{
In fact a similar calculation to (\ref{eq:DeltaSigmaExplicit}), (\ref{eq:detSigmaSchematic}) goes through for arbitrary elements of the 8-dimensional symplectic canonical group of the form $U = \exp{(\frac 12 \varphi^{MN} Z_{MN})}$
(for notation see (\ref{eq:MultimodeSqueezed})). The analogue of the $8 \times 8$ matrix $\Xi$ is 
${\varphi_K}^L$ whose trace $J^{KL}\varphi_{KL}$ vanishes by symmetry. That $\det(\Sigma)$ is in fact symplectic invariant is suggested by the form (\ref{eq:MultimodeSqueezed});
however, we do not exploit this invariance in the further discussion because it is special to the oscillator representation.} is established. 

A more general picture of invariants of the type $\sigma = \det(\Sigma)$ is as follows. Taking the generators of the quaplectic algebra
$Z_\mu$, $\overline{Z}_\nu$ and ${E^\mu}_\nu$, the symmetric algebra of all polynomials to a certain degree $K$ reduced with respect the the homogeneous $U(3,1)$ group yields a certain class of irreducible tensors of the form ${W^{\mu_1 \mu_2 \cdots \mu_k}}_{\nu_1 \nu_2 \cdots \nu_\ell}$ of contravariant/covariant rank $(k,\ell)$ (wherein both the contravariant and covariant affices have definite Young permutation symmetry type, and for which all traces vanish). Amongst these will of course be operators which commute with $U(3,1)$, corresponding to $(0,0)$ tensors. For example in an obvious notation modelled on (\ref {eq:CasimirGeneric}), at degree 1, 2 and 3 we have respectively the singlets (where $\mbox{tr}(E) \equiv {\mathcal N}$) 
\begin{align}
\mbox{tr}(E); \qquad  \mbox{tr}(E^2), \, (\mbox{tr}(E))^2, \, & \, \mbox{tr}(Z \overline{Z}); \quad \mbox{and}  \nonumber \\
            \mbox{tr}(E^3), \,  \mbox{tr}(E)\mbox{tr}(E^2), \,  (\mbox{tr}(E))^3, \,& \, \mbox{tr}(E)\mbox{tr}(Z \overline{Z}), \, \mbox{tr}(E Z \overline{Z});
\end{align}
from these and their higher-degree counterparts the Casimirs of the whole algebra are constructed as in (\ref {eq:CasimirGeneric}). However, for a fixed state $|\psi \rangle$
we can consider the classical tensors
\[
{T^{\mu_1 \mu_2 \cdots \mu_k}}_{\nu_1 \nu_2 \cdots \nu_\ell}: = 
                                         \langle \psi |{W^{\mu_1 \mu_2 \cdots \mu_k}}_{\nu_1 \nu_2 \cdots \nu_\ell} |\psi \rangle
\]
and in turn consider polynomials in these, and again look for quaplectic invariants.  
In fact for (free) oscillators, such as the scalar case in the quaplectic context (the relativistic oscillator) it is well known that such generalised correlation functions typically reduce to functions of the two-point correlator or propagator, in this case the ordinary covariance matrix $\Sigma$. Thus objects such as $\sigma = \det(\Sigma)$ are useful both because they suffice to characterise reciprocally-invariant quantum aspects of the scalar case, but also because they are the simplest in a hierarchy of such invariant quantities in general representations of the quaplectic algebra. 
\end{appendix}
\vfill
\pagebreak
%%%%%%%%%%%%%%%%%%%%%Bibliography from bibtex database here
{
% \bibliography{name}
{\small
%\bibliography{PhysGroupTh}
%\bibliographystyle{plain}
%\def\topsep{0pt}
%\def\parsep{0pt plus 5pt minus 1pt}
%\def\itemsep{-0.5ex}

}
\vfill \noindent 
{\sc 
%Authors here\\
Peter D. Jarvis, University of Tasmania, School of Mathematics and Physics,
GPO Box 252C, 7001 Hobart, TAS, Australia, 
{\small\tt Peter.Jarvis@utas.edu.au} \\
Stuart Morgan, University of Tasmania, School of Mathematics and Physics,
GPO Box 252C, 7001 Hobart, TAS, Australia, 
{\small\tt Stuart.Morgan@utas.edu.au} \\
Report Number UTAS-PHYS-2005-}
\end{document}